\documentclass[letterpaper,10pt,conference]{ieeeconf}
\IEEEoverridecommandlockouts
\usepackage{amsmath,amssymb,amsthm,mathtools}
\usepackage{booktabs,array,graphicx,microtype}
\usepackage[hidelinks]{hyperref}
\usepackage[caption=false,font=footnotesize]{subfig}
\newtheorem{theorem}{Theorem}
\newtheorem{proposition}[theorem]{Proposition}
\theoremstyle{remark}\newtheorem{remark}[theorem]{Remark}
\newcommand{\R}{\mathbb R}
\newcommand{\norm}[1]{\left\lVert#1\right\rVert}
\newcommand{\abs}[1]{\left|#1\right|}
\newcommand{\sym}{\operatorname{sym}}

\newcommand{\nz}{{n_z}}
\newcommand{\vbar}{\bar v}
\newcommand{\X}{\mathcal X}

\newcommand{\T}{\mathcal T}
\usepackage{comment}

\title{\bfseries Residual-Corrected Physics-Informed Neural Network KKL Observers with Input-to-State Stability Guarantees}
\author{Hannah Berin-Costain, Jun Liu, and Kirsten Morris%
\thanks{This work was supported in part by the Natural Sciences and
Engineering Research Council (NSERC) of Canada.}%
\thanks{The authors are with the Department of Applied Mathematics,
University of Waterloo, Waterloo, ON N2L 3G1, Canada (e-mail:
hiberinc@uwaterloo.ca; j.liu@uwaterloo.ca; kmorris@uwaterloo.ca).}}

\begin{document}
\maketitle
\thispagestyle{empty}\pagestyle{empty}

\begin{abstract}
Physics-informed neural network (PINN) Kazantzis--Kravaris--Luenberger (KKL) observers require an approximate solution to a KKL partial differential equation (PDE) and an approximate inverse of this solution. Either approximation can leave a nonzero estimation bias even without measurement noise. In this paper, we remove both errors from the implemented observer by construction. We build an encoder containing an explicitly invertible neural map. Its closed-form inverse provides the internal state reconstruction used to evaluate the learned KKL residual and correct the latent dynamics. This construction makes the synchronization manifold exactly invariant for every value of the network weights. Based on contraction of the residual-corrected observer dynamics, we prove an input-to-state stability (ISS) result for the implemented error dynamics with respect to measurement noise. A direct consequence is that the estimation error converges exponentially to zero in the noise-free setting. We provide two numerical examples, the Van der Pol oscillator and the chaotic Lorenz system, in which the residual-corrected observers achieve exact noise-free convergence and improved error responses compared with released PINN--KKL baselines.
\end{abstract}

\section{Introduction} \label{sec:intro}

A Kazantzis--Kravaris--Luenberger (KKL) observer estimates the state of a nonlinear system by simulating a
stable auxiliary system driven by the measurement. The construction rests on
a change of coordinates: an injective map 
carries the system state to a latent state whose dynamics are, by design, contracting and driven by the measured
output. Running those latent dynamics and inverting the coordinate map then
recovers the state. Classical results give broad conditions under which the
required map exists, characterize it as a solution of a first-order partial
differential equation (PDE), and guarantee a continuous left inverse
\cite{kazantzis1998,andrieu2006,pachy2024}.

Because the coordinate map and its inverse rarely have closed forms and grid-based solutions 
can be both expensive to compute and cumbersome to implement in an observer, a line
of recent work learns both with neural networks, training an encoder against the
KKL PDE in the style of physics-informed neural networks (PINNs)
and training a decoder to invert the encoder
\cite{ramos2020,niazi2026,berincostain2026,galimberti2026}. These learned
observers inherit two approximation gaps: a finite network does not satisfy
the PDE exactly, so the mismatch acts as a
persistent disturbance that stable latent dynamics attenuate but do not
remove; and the learned decoder does not invert the learned encoder exactly,
so even a perfectly synchronized latent state is decoded with an error.
Together the two gaps produce a nonzero estimation bias even without measurement
noise.

In this paper, we close both gaps in learned KKL observers by construction. 
We build an encoder containing an explicitly invertible neural map. Its
closed-form inverse provides the internal reconstruction used to evaluate
the learned KKL residual and correct the latent dynamics. This
construction makes the synchronization manifold exactly invariant for every value of the network weights. We note that 
Andrieu and Praly analyzed an approximate KKL map by feeding its equation error back through a left
inverse under an incremental small-residual condition \cite{andrieu2006}.
Our focus here is on a learning-based implementation. We also provide
an input-to-state stability (ISS) theorem for the error dynamics, which implies
exact exponential convergence of the estimation error in the noise-free setting. We highlight the following contributions:
\begin{enumerate}
\item We build the encoder with an explicitly invertible neural component, so
the residual correction is evaluated through a closed-form inverse. 
\item We reconstruct the state with a second network that is an exact left
inverse of the encoder for every value of its weights yet uses every latent
coordinate, so noise-aware training can improve estimation performance.
\item We prove an ISS theorem for the corrected dynamics under explicitly
verifiable conditions on a compact tube, with exact noise-free convergence
and an explicit ultimate error bound with respect to noise.
\item We show that the proposed construction leads to exact convergence in
the noise-free setting and improves the error response on two numerical
examples.
\end{enumerate}

Our work relates to several lines of research beyond the KKL literature
above. High-gain observers and their refinements
\cite{astolfi2015,bernard2017} and contraction-based observer designs
\cite{lohmiller1998,manchester2018,yi2021,bonnabel2015} certify
convergence from structural conditions on the system equations, whereas
the KKL route moves the structure into the latent target and must instead
certify a learned coordinate map. Within the KKL literature, gain tuning
\cite{buissonfenet2023} and set-valued extensions for indistinguishable
dynamics \cite{bernard2024} refine the design space, and neural observers
have also been constructed outside the KKL framework \cite{breiten2021}.
The exact invariance established here is a synchronization property, e.g., as considered in
the state reconstruction for the Lorenz system through a
transmitted signal \cite{cuomo1993}. Finally, the contraction-based sufficient conditions of
Theorem~\ref{thm:iss} are inequalities over compact sets, and are
therefore amenable to neural network verification
\cite{zhang2018crown,fitzsimmons2024}.

\section{Exact residual-corrected observers}
\label{sec:construction}

\subsection{System, latent dynamics, and residual}

Consider the system
\begin{equation}
 \dot x=f(x),\qquad m=h(x)+v,                               \label{eq:system}
\end{equation}
where \(x(t)\in\R^{n_x}\) is the state, \(f:\R^{n_x}\to\R^{n_x}\) is locally Lipschitz, \(h:\R^{n_x}\to\R^{n_y}\) is a continuously differentiable output map, \(m(t)\in\R^{n_y}\) is the measured signal, and \(v\) is a measurable, essentially
bounded measurement disturbance taking values in \(\R^{n_y}\). Let \(\X\subset\R^{n_x}\) be a compact set that is positively invariant for \(\dot x=f(x)\). 

A typical procedure for designing a KKL observer selects a latent dimension \(\nz\ge n_x\), finds a continuously differentiable encoder \(\T: \R^{n_x} \to \R^{\nz}\), and constructs a target vector field \(\varphi : \R^{\nz}\times\R^{n_y} \to \R^{\nz}\) such that the latent dynamics
\[
\dot z=\varphi(z,m)
\] 
contracts
in \(z\) for every fixed input signal. One common choice is \(\varphi(z,m)=Az+Bm\), where \(A\) is a Hurwitz matrix. The encoder solves the KKL equation
\[
\frac{\partial \T}{\partial x}(x) f(x) = \varphi(\T(x), h(x))
\]
on \(\X\). Solving the KKL equation analytically or numerically can be challenging. Recent works have trained physics-informed neural networks for solving it \cite{ramos2020,niazi2026,berincostain2026}. A trained encoder $\hat \T$ satisfies this equation only approximately, and we define the \emph{residual}
\begin{equation}
 r(x):=\frac{\partial \hat \T}{\partial x}(x) f(x) - \varphi(\hat \T(x), h(x)),\qquad x\in \R^{n_x}.           \label{eq:residual}
\end{equation}
Although the KKL equation is imposed only on \(\X\), every map in
\eqref{eq:residual} is defined on all of \(\R^{n_x}\), so \(r\) is as well;
this matters below, where \(r\) is evaluated at reconstructed points that
need not lie in \(\X\).

\subsection{Residual-corrected KKL observer}

With an exact KKL pair \((\T,\varphi)\), reconstruction works by inversion:
the solution of \(\dot z=\varphi(z,m)\) contracts to \(\T(x(t))\), and a
continuous left inverse of \(\T\) then recovers the state
\cite{kazantzis1998,andrieu2006}. A trained encoder with a nonzero residual
breaks the first step: \(\hat \T(x(t))\) is not a solution of
\(\dot z=\varphi(z,m)\), so the latent error does not vanish even with
\(v\equiv 0\), and the error bounds of \cite{niazi2026,berincostain2026}
therefore contain a bias term proportional to the residual bound.

If the true state \(x(t)\) were available, adding \(r(x(t))\) to the latent
dynamics would cancel the training error exactly along the system trajectory.
Since \(x(t)\) is unknown, we instead evaluate \(r\) at a state estimate
reconstructed from the latent state \(z\). For this purpose we
require a continuous map \(Q_0:\R^{\nz}\to\R^{n_x}\), called the
\emph{internal reconstruction map}, satisfying the exact identity
\begin{equation}
 Q_0(\hat \T(x))=x\qquad\text{for every }x\in \X.                  \label{eq:retraction}
\end{equation}
Given \(Q_0\), define the \emph{corrected vector field} and the
\emph{residual-corrected observer} by
\begin{align}
 F(z,m)&:=\varphi(z,m)+r(Q_0(z)),                              \label{eq:corrected-field}\\
 \dot z&=F(z,m),\qquad \hat x=Q(z),                         \label{eq:observer}
\end{align}
where the \emph{output reconstruction} \(Q:\R^{\nz}\to\R^{n_x}\) is a second
continuous map that is also exact on the encoder image \(\hat\T(\X)\), that is,
\begin{equation}\label{eq:output-exact}
    Q(\hat \T(x))=x,
    \qquad x\in\mathcal X,
\end{equation} 
but need not equal \(Q_0\). The reason for keeping
two maps is explained in Section~\ref{sec:two-maps}.

\begin{proposition}[Exact synchronization manifold]\label{prop:exact-graph}
    Suppose \eqref{eq:retraction} holds. Then for every \(x\in \X\),
    \begin{equation}
     F(\hat \T(x),h(x))=\frac{\partial \hat \T}{\partial x}(x)f(x).        \label{eq:graph-identity}
    \end{equation}
    Consequently, if \(x(t)\) solves \(\dot x=f(x)\) in \(\X\) and \(v\equiv0\),
    then \(z(t)=\hat \T(x(t))\) solves \(\dot z=F(z,h(x(t)))\). Along this solution, \eqref{eq:output-exact} gives $\hat x(t) = x(t)$.
\end{proposition}

\begin{proof}
    Using \eqref{eq:corrected-field} and the definition of the residual,
    \begin{align*}
        F(\hat \T(x),h(x)) &= \varphi(\hat \T(x),h(x))+r(Q_0(\hat \T(x))) \\
        & = \varphi(\hat \T(x),h(x))+r(x) \\
        & = \frac{\partial \hat \T}{\partial x}(x) f(x).
    \end{align*}
    The chain rule therefore gives
    \[
    \frac{d}{dt}\hat \T(x(t)) = F(\hat \T(x(t)),h(x(t))).
    \]
    Thus, $z(t) = \hat \T(x(t))$ satisfies the noise-free observer equation. The reconstruction identity gives $\hat x (t) = Q(\hat \T(x(t))) = x(t)$.
\end{proof}

The correction makes the encoded system trajectory an exact solution of the observer dynamics, regardless of the residual magnitude. Convergence to this trajectory, however, is not automatic. We propose in Section~\ref{sec:theory} a contraction condition on the corrected vector field $F$ and derive an ISS bound for the estimation error, which reduces to exact exponential convergence when $v \equiv 0$. 

\subsection{Internal and output reconstruction}\label{sec:two-maps}

Both reconstruction maps $Q_0$ and $Q$ are exact on the learned encoder image:
\[
Q_0(\hat \T(x)) = Q(\hat \T(x)) = x, \qquad x \in \X.
\]
However, they serve different purposes away from the image $\hat \T (\X)$. 

The internal reconstruction map $Q_0$ is used to evaluate the residual
correction $r(Q_0(z))$ in \eqref{eq:corrected-field} and therefore directly
affects the observer dynamics. In particular, the contraction condition in Section~\ref{sec:theory} depends on the derivatives of $r \circ Q_0$. For this reason, $Q_0$ is chosen to be simple and fixed. In the proposed architecture, the first \(n_x\) components of the encoder form an invertible map \(T_0\). The internal reconstruction is defined explicitly by \(Q_0(z)=T_0^{-1}(z_{1:n_x})\). This closed form also allows the correction to be evaluated without solving a nonlinear equation online.

The output reconstruction $Q$ is used only to compute $\hat x = Q(z)$ and does not enter the latent dynamics. It may therefore use all \(\nz\) latent coordinates and can be trained to reduce the reconstruction error under measurement noise. Its architecture guarantees $Q(\hat \T(x)) = x$ for every $x$ and every value of the trainable parameters. Thus, training $Q$ cannot introduce an error when $z = \hat \T(x)$ and does not change the contraction properties of the corrected field. Its sensitivity in a tube around the encoder image affects the physical-state error bound in Section~\ref{sec:theory}.

Separating the two maps therefore separates the stability condition from the training of the
observer output. The contraction condition involves the fixed map $Q_0$, while the reconstruction
bound involves $Q$. The simpler choice $Q = Q_0$ is also allowed.

\section{Contraction-based ISS analysis}
\label{sec:theory}

We present the main theoretical results on contraction and ISS analysis of the residual-corrected observer architecture in this section. 

\subsection{Notation and standing assumptions}

Let \(P\in\R^{\nz\times\nz}\) be a symmetric positive definite matrix, fixed
throughout this section. It defines the inner product
\(\langle a,b\rangle_P=a^\top Pb\) and the norm
\(\norm{e}_P=(e^\top Pe)^{1/2}\) on \(\R^{\nz}\); \(\norm{\cdot}_2\) denotes
the Euclidean norm on \(\R^{n_x}\) and \(\R^{n_y}\) and, for matrices, the induced spectral norm.
For a square matrix \(M\), \(\sym M=(M+M^\top)/2\), and \(M\preceq M'\) means
that \(M'-M\) is positive semidefinite.

For a tube radius \(R>0\), define
\begin{equation}
 {\cal E}_R:=\bigl\{(x,z)\in \X\times\R^{\nz}:
 \norm{z-\hat \T(x)}_P\le R\bigr\},                               \label{eq:tube}
\end{equation}
the set of system--latent state pairs whose latent error, measured in the
\(P\)-norm, does not exceed \(R\). For a noise bound \(\vbar>0\), the
measured signal ranges over
\begin{equation}
 {\cal M}_{\vbar}:=\{h(x)+w:x\in \X,\ \norm{w}_2\le\vbar\},     \label{eq:meas-set}
\end{equation}
which is a compact set. All disturbances \(v\) are measurable with
\(\norm{v}_\infty\le\vbar\), where
\(\norm{v}_{\infty,[0,t]}=\operatorname*{ess\,sup}_{s\in[0,t]}\norm{v(s)}_2\)
and \(\norm{v}_\infty=\norm{v}_{\infty,[0,\infty)}\). Since \(F\) in
\eqref{eq:corrected-field} is continuous and locally Lipschitz in \(z\)
(being built from continuously differentiable maps), the observer equation
with a measurable input admits unique absolutely continuous
(Carath\'eodory) solutions \cite[Appendix~C]{sontag1998}.

\subsection{The ISS theorem}

We now state the main theoretical result of the paper. 

\begin{theorem}[ISS of the residual-corrected observer]
\label{thm:iss}
Let \(\hat \T\) be continuously differentiable on a neighborhood of \(\X\),
and let \(Q_0\) satisfy \eqref{eq:retraction}. Fix \(R>0\) and
\(\vbar>0\), and suppose that \(F\) is continuously differentiable in \(z\)
and \(Q\) is continuously differentiable, on an open neighborhood of the
latent projection \(\{z\in\R^{\nz}:(x,z)\in{\cal E}_R\ \text{for some }x\in\X\}\)
of the tube, uniformly over \(m\in{\cal M}_{\vbar}\). Assume that \(Q\) satisfies \eqref{eq:output-exact} and that
there exist a constant \(\alpha>0\), a constant \(L_Q>0\), and a
nondecreasing function \(\Delta:[0,\vbar]\to[0,\infty)\) with
\(\Delta(0)=0\), such that the following three conditions hold: 
\begin{enumerate}
\item\textbf{(Tube contraction)} For every \(x\in\X\), every \(e\in\R^{\nz}\)
with \(\norm{e}_P\le R\), every \(s\in[0,1]\), and every
\(m\in{\cal M}_{\vbar}\), the Jacobian
\(A(x,e,s,m):= \frac{\partial F}{\partial z} \bigl(\hat \T(x)+se,\,m\bigr)\) satisfies
\begin{equation}
 A(x,e,s,m)^\top P+PA(x,e,s,m)\preceq-2\alpha P.           \label{eq:field-contract}
\end{equation}
\item\textbf{(Noise forcing bound)} For every \(x\in\X\) and every \(w\in\R^{n_y}\) with
\(\norm{w}_2\le\vbar\),
\begin{equation}
 \bigl\lVert F\bigl(\hat \T(x),h(x)+w\bigr)-F\bigl(\hat \T(x),h(x)\bigr)\bigr\rVert_P
 \le\Delta(\norm{w}_2).                                        \label{eq:forcing}
\end{equation}
\item\textbf{(Reconstruction sensitivity)} For every \(x\in\X\) and every
\(e\) with \(\norm{e}_P\le R\),
\begin{equation}
    \norm{
        \frac{\partial Q}{\partial z}
        \bigl(\hat\T(x)+e\bigr)P^{-1/2}
    }_2
    \le L_Q.
    \label{eq:decoder-gain}
\end{equation}
\end{enumerate}
Finally, assume the tube-invariance condition
\begin{equation}
 \Delta(\vbar)<\alpha R.                                   \label{eq:residence}
\end{equation}
Then, for every system trajectory \(x(\cdot)\) in \(\X\), every measurable
disturbance with \(\norm{v}_\infty\le\vbar\), and every initial latent state
with \(\norm{z(0)-\hat \T(x(0))}_P\le R\):
\begin{enumerate}
\item the tube is robustly positively invariant, that is,
\(\norm{z(t)-\hat \T(x(t))}_P\le R\) for all \(t\ge0\);
\item the latent error obeys, for all \(t\ge0\),
\begin{multline}
 \norm{z(t)-\hat \T(x(t))}_P
 \le e^{-\alpha t}\norm{z(0)-\hat \T(x(0))}_P \\
 +\frac{\Delta\bigl(\norm{v}_{\infty,[0,t]}\bigr)}{\alpha}
       \bigl(1-e^{-\alpha t}\bigr);                        \label{eq:latent-bound}
\end{multline}
\item the state estimate obeys, for all \(t\ge0\),
\begin{multline}
 \norm{\hat x(t)-x(t)}_2
 \le L_Q\,e^{-\alpha t}\norm{z(0)-\hat \T(x(0))}_P \\
 +\frac{L_Q}{\alpha}
 \Delta\bigl(\norm{v}_{\infty,[0,t]}\bigr)
 \bigl(1-e^{-\alpha t}\bigr).                              \label{eq:state-bound}
\end{multline}
\end{enumerate}
In particular, \(v\equiv0\) gives exact exponential convergence of the
estimate, and \(\norm{v}_\infty\le\vbar\) gives the ultimate bound
\begin{equation}
 \limsup_{t\to\infty}\norm{\hat x(t)-x(t)}_2
 \le \frac{L_Q\,\Delta(\vbar)}{\alpha}.                    \label{eq:ultimate}
\end{equation}
\end{theorem}

\begin{proof}
Write \(e(t):=z(t)-\hat \T(x(t))\) and \(m(t)=h(x(t))+v(t)\). The proof proceeds
in five steps.

\emph{Step 1 (error dynamics).} Along the system trajectory,
\(\frac{d}{dt}\hat\T(x(t)) = \frac{\partial\hat\T}{\partial x}(x(t))f(x(t))\), which by
Proposition~\ref{prop:exact-graph} equals \(F(\hat \T(x(t)),h(x(t)))\). Hence, for
almost every \(t\),
\begin{align}
 \dot e={}&F\bigl(\hat \T(x)+e,m\bigr)-F\bigl(\hat \T(x),m\bigr)\notag\\
 &+F\bigl(\hat \T(x),m\bigr)-F\bigl(\hat \T(x),h(x)\bigr),             \label{eq:error-split}
\end{align}
where the first difference compares the corrected field at the true and
encoded latent states under the same measured input, and the second isolates
the effect of the disturbance.

\emph{Step 2 (differential inequality for \(\norm{e}_P\)).} Suppose
\(\norm{e(t)}_P\le R\), so that the segment \(\hat \T(x)+se\), \(s\in[0,1]\), lies
in the region covered by \eqref{eq:field-contract}. By the fundamental theorem of
calculus applied to \(s\mapsto F(\hat \T(x)+se,m)\),
\begin{equation}
 F\bigl(\hat \T(x)+e,m\bigr)-F\bigl(\hat \T(x),m\bigr)
 =\Bigl(\int_0^1 A(x,e,s,m)\,ds\Bigr)e.                    \label{eq:mean-value}
\end{equation}
Denote the second difference in \eqref{eq:error-split} by \(g(t)\); by
\eqref{eq:forcing} and monotonicity of \(\Delta\),
\(\norm{g(t)}_P\le\Delta(\norm{v(t)}_2)\le\Delta(\norm{v}_{\infty,[0,t]})\).
Differentiating \(\norm{e}_P^2=e^\top Pe\) and substituting
\eqref{eq:error-split}--\eqref{eq:mean-value} gives, for almost every \(t\),
\begin{equation}
 \tfrac{d}{dt}\norm{e}_P^2
 =2\int_0^1 e^\top PA(x,e,s,m)\,e\,ds+2\,e^\top Pg.         \label{eq:vdot}
\end{equation}
For each \(s\), \eqref{eq:field-contract} gives
\(2e^\top PA(x,e,s,m)e=e^\top(A^\top P+PA)e\le-2\alpha\,e^\top Pe\).
The Cauchy--Schwarz inequality gives
\(\abs{e^\top Pg}=\abs{e^\top P^{1/2}P^{1/2}g}\le\norm{e}_P\norm{g}_P\). Therefore
\begin{equation}
 \tfrac{d}{dt}\norm{e}_P^2
 \le-2\alpha\norm{e}_P^2+2\norm{e}_P\,\Delta\bigl(\norm{v(t)}_2\bigr).
                                                            \label{eq:vdot-bound}
\end{equation}
Wherever \(e(t)\ne0\), dividing by \(2\norm{e}_P\) yields
\(\frac{d}{dt}\norm{e}_P\le-\alpha\norm{e}_P+\Delta(\norm{v(t)}_2)\). At a time
with \(e(t)=0\), the upper right Dini derivative satisfies
\(D^+\norm{e}_P\le\norm{\dot e}_P=\norm{g}_P\le\Delta(\norm{v(t)}_2)\), which is
the same inequality. Hence, in the sense of Dini derivatives and for almost
every \(t\) while the trajectory remains in the tube,
\begin{equation}
 D^+\norm{e}_P\le-\alpha\norm{e}_P+\Delta\bigl(\norm{v(t)}_2\bigr). \label{eq:scalar}
\end{equation}

\emph{Step 3 (comparison).} Let
\(\delta:=\Delta(\norm{v}_{\infty,[0,t]})\). The scalar comparison lemma
\cite[Lem.~3.4]{khalil2002} applied to \eqref{eq:scalar} with the comparison
equation \(\dot u=-\alpha u+\delta\), whose solution is
\(u(t)=e^{-\alpha t}u(0)+(\delta/\alpha)(1-e^{-\alpha t})\), gives
\eqref{eq:latent-bound} for as long as the trajectory remains in the tube.

\emph{Step 4 (robust invariance of the tube).} The system cannot leave \(\X\),
and the measured signal always lies in \({\cal M}_{\vbar}\); only the latent
error could exit. Suppose, for contradiction, that the error leaves the
tube, and let
\(t_1:=\inf\{t\ge0:\norm{e(t)}_P>R\}<\infty\). By continuity,
\(\norm{e(t_1)}_P=R\), and \eqref{eq:scalar} holds at \(t_1\). The
tube-invariance condition \eqref{eq:residence} then gives
\[
 D^+\norm{e}_P\big|_{t_1}\le-\alpha R+\Delta(\vbar)<0,
\]
so there exists \(\eta>0\) with \(\norm{e(t)}_P<R\) for
\(t\in(t_1,t_1+\eta)\). This contradicts the definition of \(t_1\) as the
infimum of exit times. Hence the error never exits, part~1) holds, and the
comparison estimate \eqref{eq:latent-bound} holds for all \(t\ge0\).

\emph{Step 5 (from the latent error to the state estimate).} Fix \(t\), write
\(e=e(t)\), and consider \(s\mapsto Q(\hat \T(x)+se)\) on \([0,1]\); the segment
lies in the region covered by \eqref{eq:decoder-gain}. Using \(Q(\hat \T(x))=x\) and the
fundamental theorem of calculus,
\[
    \hat x-x
    =
    Q\bigl(\hat\T(x)+e\bigr)-Q\bigl(\hat\T(x)\bigr)
    =
    \int_0^1
    \frac{\partial Q}{\partial z}
    \bigl(\hat\T(x)+se\bigr)e\,ds.
\]
Insert \(P^{-1/2}P^{1/2}\) and bound each factor:
\begin{align*}
\norm{\hat x-x}_2
&\le
\int_0^1
\norm{
    \frac{\partial Q}{\partial z}(\hat\T(x)+se)P^{-1/2}
}_2
\norm{P^{1/2}e}_2\,ds\\
&\le L_Q\norm{e}_P.
\end{align*}
Combining with
\eqref{eq:latent-bound} proves \eqref{eq:state-bound}. Letting
\(t\to\infty\) in \eqref{eq:state-bound} with
\(\norm{v}_\infty\le\vbar\) proves \eqref{eq:ultimate}, and setting
\(v\equiv0\) (so \(\Delta=0\)) gives exact exponential convergence.
\end{proof}

\begin{remark}[Computing the noise bound \(\Delta\)]
\label{rem:forcing}
The correction term in \eqref{eq:corrected-field} is evaluated at
\(Q_0(z)\) and has no measurement argument, so the disturbance enters
\(F\) only through the target:
\begin{align*}
 &F\bigl(\hat \T(x),h(x)+w\bigr)-F\bigl(\hat \T(x),h(x)\bigr)\\
 &\qquad=\varphi\bigl(\hat \T(x),h(x)+w\bigr)-\varphi\bigl(\hat \T(x),h(x)\bigr).
\end{align*}
We therefore obtain \(\Delta\) directly from the formula for
\(\varphi\); neither the reconstruction maps nor the residual correction enters the noise channel. For \(M\in\R^{\nz\times n_y}\), write
\(\norm{M}_P:=\norm{P^{1/2}M}_2\). If the measurement enters the target
affinely, \(\varphi(z,m)=\varphi_0(z)+Bm\), the target difference is
\(Bw\) and \(\Delta(s)=s\,\norm{B}_P\) is the smallest valid choice.
In general, if
\(\norm{ \frac{\partial\varphi}{\partial m}(z,m)}_P\le L_m\) on
\(\hat \T(\X)\times{\cal M}_{\vbar}\), integrating \(\partial\varphi/\partial m\) along
the segment from \(h(x)\) to \(h(x)+w\) gives \eqref{eq:forcing} with
\(\Delta(s)=L_m\,s\).
\end{remark}

\section{Neural realization and training} \label{sec:training}

In this section, we define the neural parameterizations of the generic maps introduced in
Section~\ref{sec:construction} and describe the training procedure. 

\subsection{Exact neural reconstruction}

Let $D_x\in\mathbb R^{n_x\times n_x}$ be diagonal and nonsingular, let $c_x\in\mathbb R^{n_x}$, and define $\bar x=D_x^{-1}(x-c_x)$. We take $c_x$ to be the midpoint of the coordinatewise training box and $D_x$ to collect its half-widths, so that $\bar x$ ranges over $[-1,1]^{n_x}$; both are fixed before training.

The first $n_x$ encoder coordinates are produced by an invertible neural map $S_\theta:\mathbb R^{n_x}\to\mathbb R^{n_x}$ constructed from affine coupling layers \cite{dinh2017}. For a binary mask $a\in\{0,1\}^{n_x}$, one layer is
\begin{equation}\label{eq:coupling}
\mathcal C(u)
=
a\odot u
+(1-a)\odot
\left[
    u\odot\exp\bigl(\ell(a\odot u)\bigr)
    +t(a\odot u)
\right],
\end{equation}
where $\odot$ is the componentwise product and $\ell$ and $t$ are
continuously differentiable feedforward networks. Because
$\exp(\ell(\cdot))$ is componentwise positive, each layer has an explicit
inverse. We take $\ell=\ell_{\max}\tanh(\widetilde\ell)$, which confines
the componentwise scaling factors to $[e^{-\ell_{\max}},e^{\ell_{\max}}]$
and keeps $S_\theta$ and its inverse well conditioned. Composing layers
with alternating masks gives a global $C^1$ diffeomorphism $S_\theta$.

A feedforward network $\tau_\theta:\mathbb R^{n_x}\to \mathbb R^{n_z-n_x}$ generates the remaining latent coordinates. The complete neural encoder is
\begin{equation}\label{eq:encoder}
    \hat  \T_\theta(x) =
    \begin{bmatrix}
        s\\
        \tau_\theta(s)
    \end{bmatrix},
    \qquad
    s=S_\theta(\bar x).
\end{equation}
Partition $z=(z_c,z_r)$ accordingly. The internal reconstruction is given by
\begin{equation}\label{eq:core-inverse}
    Q_0(z)=c_x+D_xS_\theta^{-1}(z_c).
\end{equation}

To construct an output map that uses all latent coordinates, define the consistency error
\begin{equation}\label{eq:defect}
    \eta(z)=z_r-\tau_\theta(z_c)
\end{equation}
and let $M_\psi: \mathbb R^{n_x}\times\mathbb R^{n_z-n_x} \to\mathbb R^{n_x\times(n_z-n_x)}$ be a feedforward network. The output reconstruction is given by
\begin{equation}\label{eq:exact-decoder}
    Q_\psi(z) = c_x+D_xS_\theta^{-1}
    \left(
        z_c+M_\psi(z_c,\eta(z))\eta(z)
    \right).
\end{equation}

\begin{proposition}[Exact reconstruction] \label{prop:decoder-exact}
    For every value of $\theta$ and $\psi$, $Q_0(\hat \T_\theta(x)) = Q_\psi(\hat \T_\theta(x)) =x$ for all $x\in\mathbb R^{n_x}$.
\end{proposition}

\begin{proof}
    For $z=\hat \T_\theta(x)$, $z_c=S_\theta(\bar x)$, $z_r=\tau_\theta(z_c)$ so $\eta(z)=0$. Substitution into
    \eqref{eq:core-inverse} and \eqref{eq:exact-decoder} gives
    \[
        Q_0(z)=Q_\psi(z) =c_x+D_xS_\theta^{-1}(S_\theta(\bar x)) =x.
    \]
\end{proof}

Therefore, both reconstruction identities hold for all choices of parameter values. The map $Q_0$ is used in the corrected dynamics, while $Q_\psi$ uses all latent coordinates to produce the state estimate.

\subsection{Contracting target field}

For the nonlinear latent dynamics, we use
\begin{equation}\label{eq:contracting-target}
    \begin{aligned}
        \varphi_\theta(z,m)
        ={}&
        \frac{1}{2}(K-K^\top)z-cz+b+Bm\\
        &-
        W^\top\operatorname{diag}(q)
        \tanh(Wz+Um+d),
    \end{aligned}
\end{equation}
where $K\in\R^{\nz\times\nz}$, $b\in\R^{\nz}$, $B\in\R^{\nz\times n_y}$,
$W\in\R^{n_\varphi\times\nz}$, $U\in\R^{n_\varphi\times n_y}$, and
$d,q\in\R^{n_\varphi}$ are trainable, $c>0$ is fixed, and the entries of
$q$ are positive.

\begin{proposition}[Structural contraction of the target field]\label{prop:target-contraction}
    For every parameter value and every $(z,m)$,
    \[
        \operatorname{sym}
        \left(
            \frac{\partial\varphi_\theta}{\partial z}(z,m)
        \right)
        \preceq-cI_{n_z}.
    \]
\end{proposition}

\begin{proof}
    Differentiating \eqref{eq:contracting-target} gives
    \[
        \frac{\partial\varphi_\theta}{\partial z}
        =
        \frac{1}{2}(K-K^\top)-cI_{n_z}
        -W^\top\Gamma(z,m)W,
    \]
    where
    \[
        \Gamma(z,m)
        =
        \operatorname{diag}
        \left(
            q\odot\operatorname{sech}^2(Wz+Um+d)
        \right)
        \succeq0.
    \]
    The first term is skew-symmetric, and hence
    \[
        \operatorname{sym}
        \left(
            \frac{\partial\varphi_\theta}{\partial z}
        \right)
        =
        -cI_{n_z}-W^\top\Gamma(z,m)W
        \preceq-cI_{n_z}.
    \]
\end{proof}

\begin{remark}[Targets used in the numerical examples]
\label{rem:example-targets}
In Section~\ref{sec:numerics}, the Lorenz observer uses the nonlinear
target \eqref{eq:contracting-target}. By
Remark~\ref{rem:forcing}, its noise bound \(\Delta(s)=L_m\,s\) only
requires \(\norm{\frac{\partial\varphi_\theta}{\partial m}(z,m)}_P\le L_m\); differentiating
\eqref{eq:contracting-target} in \(m\) and using
\(\operatorname{sech}^2\le1\) gives the global choice
\begin{equation}\label{eq:closed-input-gain}
    L_m = \|P^{1/2}B\|_2 +
    \left\|
        P^{1/2}W^\top\operatorname{diag}(q)
    \right\|_2
    \|U\|_2.
\end{equation}
The Van der Pol observer instead uses the linear target
\(\varphi(z,m)=Az+Bm\), with \(A\) diagonal and Hurwitz taken from the
released PINN--KKL inference poles, which matches the target dynamics of
the baseline comparison; its noise bound is exactly
\(\Delta(s)=s\,\norm{B}_P\).
\end{remark}

\subsection{Two-stage training}

In the first stage, the parameters of $\hat \T_\theta$ and $\varphi_\theta$ are trained jointly.
Since $Q_0$ is defined through $S_\theta^{-1}$, it changes with the encoder. The loss function for this stage is given by
\begin{equation}\label{eq:observer-loss}
    \begin{aligned}
        \mathcal L_{\mathrm{obs}}
        ={}&
        \lambda_r\mathbb E_x\|r_\theta(x)\|_2^2
        +\lambda_c\mathcal L_{\mathrm{con}}
        +\lambda_h\mathcal L_{\mathrm{hard}}\\
        &+\lambda_i\mathcal L_{\mathrm{inv}}
        +\lambda_g\mathcal L_{\mathrm{gain}}.
    \end{aligned}
\end{equation}
The first term penalizes the KKL residual \eqref{eq:residual}. With a
smoothed hinge \(\rho\) and a margin \(c_0>0\), the contraction term
samples the one-sided rates of \eqref{eq:field-contract}:
\begin{equation}\label{eq:contraction-loss}
 \mathcal L_{\mathrm{con}}
 =\max_i\; \rho \Bigl(\tfrac1\varepsilon\,d_i^\top
 \bigl[F(\zeta_i+\varepsilon d_i,m_i)-F(\zeta_i,m_i)\bigr]+c_0\Bigr),
\end{equation}
where \(\zeta_i\) lies in a small tube around \(\hat \T_\theta(x_i)\),
\(d_i\) is a random unit direction, and \(m_i=h(x_i)\). The term
\(\mathcal L_{\mathrm{hard}}\) applies the same hinge to
\(\lambda_{\max}\left(\sym\frac{\partial F}{\partial z}\right)\) at the worst states encountered
during training, which are retained rather than resampled;
\(\mathcal L_{\mathrm{inv}}\) hinges the sampled amplification
\(\norm{Q_0(z+\delta)-Q_0(z)}_2/\norm{\delta}_2\) of the internal
reconstruction \eqref{eq:core-inverse}; and \(\mathcal L_{\mathrm{gain}}\)
squares the Euclidean analogue of the measurement gain
\eqref{eq:closed-input-gain}. 

In the second stage, $\hat \T_\theta$, $\varphi_\theta$, and $Q_0$ are frozen. We simulate noisy
observer trajectories and train only $Q_\psi$. For recorded pairs $(z_i,x_i)$ with noise amplitudes $\bar v_i$, the loss is
\begin{equation}\label{eq:output-loss}
    \mathcal L_{\mathrm{out}}
    =
    \frac{1}{M}
    \sum_{i=1}^{M}
    \frac{
        \left\|
            D_x^{-1}\bigl(Q_\psi(z_i)-x_i\bigr)
        \right\|_2^2
    }{
        \max\{\bar v_i,\bar v_{\min}\}^2
    },
\end{equation}
where $\bar v_{\min}>0$ handles noise-free records. Proposition~\ref{prop:decoder-exact} remains valid throughout
this stage. Since $Q_\psi$ does not enter $F$, output training does not change the latent dynamics or their contraction properties.

Table~\ref{tab:architecture} lists the frozen dimensions and optimization sizes for both systems; the functional construction is identical, and only widths and the target family differ.

\begin{table}[t]
\vspace{5pt}
\caption{Architecture and training configuration}
\label{tab:architecture}
\centering
\setlength{\tabcolsep}{3.2pt}
\begin{tabular}{lcc}
\toprule
Quantity & Van der Pol & Lorenz\\
\midrule
Physical / latent dimension \(n_x/\nz\) & \(2/5\) & \(3/7\)\\
Coupling layers / width & \(4/16\) & \(4/20\)\\
$\tau_\theta$ network width & 24 & 32\\
Target family & linear & nonlinear \eqref{eq:contracting-target}\\
Target contraction rate \(c\) & 1.5 & 1.5\\
Observer update steps & 500 & 300 (refine)\\
Observer batch size & 64 & 32\\
Output-head width & 64 & 64\\
Output-head steps & 4000 & 3000\\
Noisy training records \(M\) & 102400 & 57600\\
\bottomrule
\end{tabular}
\end{table}

\section{Numerical studies}\label{sec:numerics}

We evaluate the residual-corrected observers on the Van der Pol (VDP) and Lorenz systems, and compare them with the released PINN--KKL checkpoints of Niazi et al. \cite{niazi2026}. Our observer uses the residual-corrected field \(F(z,m)=\varphi_\theta(z,m)+r_\theta(Q_0(z))\) and the trained exact output map \(Q_\psi\). 

Each system is simulated for 40 seconds from six fixed initial states. Three are selected directly, while the remaining three are obtained by evolving the system from those states for a fixed burn-in period. Both observers are evaluated using the same system trajectories, measurement signals, and initial state estimates. The initial estimation error has Euclidean norm \(0.2\) for the Van der Pol system and \(2\) for the Lorenz system.

For each trajectory \(j\), we compare noise-free measurements with the bounded deterministic signal
\begin{equation}
     v_j(t)=\bar v\bigl[0.6\sin(1.9t+\theta_j)
            +0.4\sin(5.7t+0.23+\theta_j)\bigr],
     \label{eq:comparison-noise}
\end{equation}
where the six phases are equally spaced from \(0.17\) to \(5.31\). 

All trajectories are integrated in double precision with the adaptive Dormand--Prince integrator of order 8 (DOP853) \cite{hairer1993} and sampled every \(0.01\) seconds. Figure~\ref{fig:baseline-comparisons} shows maximum Euclidean state-estimation errors across the six trajectories at each time. Under noise-free measurements, the corrected observer's error saturates at the accuracy limit of the numerical integration, which decreases under stricter integration tolerances. Tables~\ref{tab:vdp}--\ref{tab:lorenz} report the overall RMSE computed from all trajectories and sampled times in \([30,40]\), together with the largest error over the same interval.

\subsection{Van der Pol}

Consider the Van der Pol oscillator
\begin{equation}
 \dot x_1=x_2,\qquad
 \dot x_2=3(1-x_1^2)x_2-x_1,\qquad h(x)=x_1.                \label{eq:vdp}
\end{equation}
The observer uses the architecture of Section~IV with the configuration in Table~\ref{tab:architecture}. Its linear latent target is
\begin{align*}
    \varphi_\theta(z,m)&=Az+Bm,\\
    A &=\operatorname{diag}(-1.5,-2.25,-3,-3.75,-4.5).
\end{align*}
The poles match those used by the released baseline \cite{niazi2026}. The output map is trained using noise amplitudes \(0.01\) and \(0.1\).

Table~\ref{tab:vdp} and Fig.~\ref{fig:baseline-comparisons}(a) show that the corrected
observer reaches the integration accuracy limit without noise, while the
released baseline retains a nonzero error. At \(\bar v=10^{-4}\), the
corrected observer has approximately \(8.19\) times smaller overall RMSE
and \(4.91\) times smaller maximum error on the six test trajectories.

\begin{table}[t]
\vspace{5pt}
\caption{VDP estimation errors on time interval \([30,40]\): overall RMSE and maximum across all six trajectories.}
\label{tab:vdp}
\centering
\begin{tabular*}{\columnwidth}{
    @{\extracolsep{\fill}}lcc@{}
}
\toprule
\(\bar v\) & Proposed observer & Released PINN--KKL \cite{niazi2026}\\
 & RMSE\,/\,maximum & RMSE\,/\,maximum\\
\midrule
\(0\) &
\(5.47\!\times\!10^{-11}\,/\,9.53\!\times\!10^{-10}\) &
\(7.36\!\times\!10^{-4}\,/\,1.90\!\times\!10^{-3}\)\\
\(10^{-4}\) &
\(9.12\!\times\!10^{-5}\,/\,4.07\!\times\!10^{-4}\) &
\(7.47\!\times\!10^{-4}\,/\,2.00\!\times\!10^{-3}\)\\
\(10^{-3}\) &
\(9.12\!\times\!10^{-4}\,/\,4.07\!\times\!10^{-3}\) &
\(1.19\!\times\!10^{-3}\,/\,4.19\!\times\!10^{-3}\)\\
\bottomrule
\end{tabular*}
\end{table}

\subsection{Lorenz}

Consider the chaotic Lorenz system
\begin{align}
 \dot x_1&=10(x_2-x_1),\qquad
 \dot x_2=x_1(28-x_3)-x_2, \notag\\
 \dot x_3&=x_1x_2-\tfrac83x_3,\qquad h(x)=x_2.              \label{eq:lorenz}
\end{align}
The observer uses the architecture of Section~IV with the configuration in Table~\ref{tab:architecture}. The nonlinear
latent target is given by \eqref{eq:contracting-target}. The released PINN--KKL baseline uses a linear latent target and the same measured state \(x_2\).

\begin{figure*}[t]
    \centering

    \subfloat[Van der Pol\label{fig:vdp-baseline}]{
        \includegraphics[width=0.985\textwidth]
        {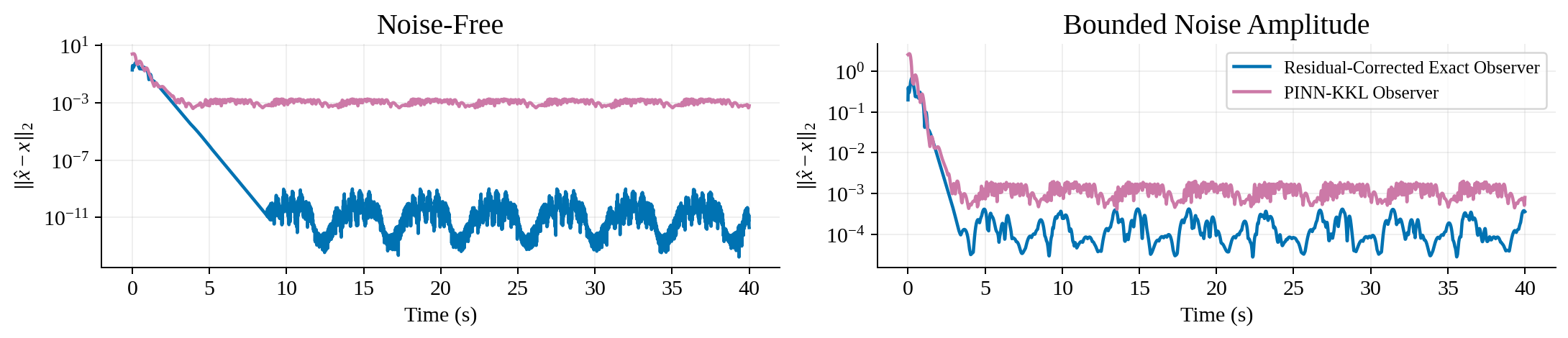}
    }

    \subfloat[Lorenz\label{fig:lorenz-baseline}]{
        \includegraphics[width=0.985\textwidth]
        {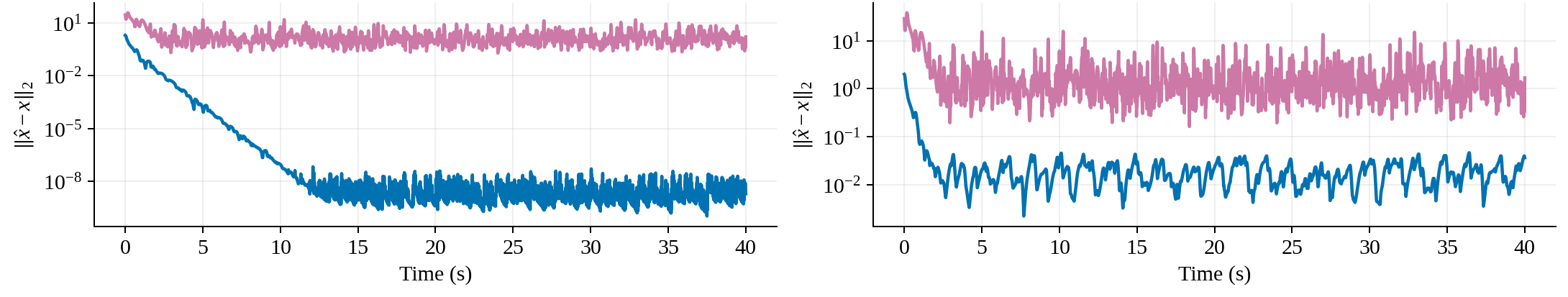}
    }

    \caption{State-estimation error for the proposed
    residual-corrected observer and the released PINN--KKL
    observer of Niazi et al.~\cite{niazi2026}. Each curve shows the maximum Euclidean error across the
    corresponding test trajectories. The left panels use
    noise-free measurements, while the right panels use bounded
    measurement noise with amplitude \(10^{-4}\) for the Van der
    Pol system and \(0.01\) for the Lorenz system. Each comparison
    uses the same system trajectories and measurement signals for
    both observers.}
    \label{fig:baseline-comparisons}
\end{figure*}

Table~\ref{tab:lorenz} and Fig.~\ref{fig:baseline-comparisons}(b) show substantially smaller errors for the corrected observer on these six trajectories. Without noise its tail error reaches the integration accuracy limit; at \(\bar v=10^{-2}\), its overall RMSE is approximately \(93.8\) times smaller than that of the released checkpoint. At \(\bar v=10^{-1}\), its overall RMSE is approximately \(9.61\) times smaller than that of the released checkpoint.

\begin{table}[t]
\vspace{5pt}
\caption{Lorenz estimation errors on time interval \([30,40]\): overall RMSE
and maximum across all six trajectories.}
\label{tab:lorenz}
\centering
\begin{tabular*}{\columnwidth}{
    @{\extracolsep{\fill}}lcc@{}
}
\toprule
\(\bar v\) & Proposed observer & Released PINN--KKL \cite{niazi2026}\\
 & RMSE\,/\,maximum & RMSE\,/\,maximum\\
\midrule
\(0\)       & \(2.69\!\times\!10^{-9}\,/\,4.61\!\times\!10^{-8}\)
            & \(1.20\,/\,15.1\)\\
\(10^{-2}\) & \(0.0128\,/\,0.0463\)
            & \(1.20\,/\,15.1\)\\
\(10^{-1}\) & \(0.128\,/\,0.463\)
            & \(1.23\,/\,14.3\)\\
\bottomrule
\end{tabular*}
\end{table}

\subsection{Numerical assessment of the stability conditions}

After training, each observer is frozen and evaluated against the conditions of Theorem~\ref{thm:iss}. We first fit a
constant metric $P\succ0$ using corrected-field Jacobians on the encoder image. Independent system states, latent perturbations, and measurement offsets are sampled throughout each candidate tube. At every sample, we evaluate the Jacobian of the corrected field $F(z,m)=\varphi_\theta(z,m) + r_\theta(Q_0(z))$, rather than the target field \(\varphi_\theta\) alone. We also estimate the measurement forcing and output sensitivity. A tube is retained only when the sampled contraction and tube-invariance conditions of Theorem~\ref{thm:iss} are satisfied.

The screen gives contraction rates \(0.173\) on the VDP tube \(R=0.003\) and \(0.228\) on the Lorenz tube \(R=0.015\). Both tubes pass the sampled invariance test at the principal noise levels, \(\bar v=10^{-4}\) and \(10^{-2}\), respectively. These results are sampled numerical evidence for the conditions of Theorem~\ref{thm:iss}, not continuous-domain certificates. Establishing the same inequalities uniformly over the tubes, with neural network verification \cite{zhang2018crown,fitzsimmons2024} or rigorous numerical enclosures \cite{kapela2021}, is a natural direction for future work.

\section{Conclusion}\label{sec:conclusion}

In this work we have developed a neural KKL observer that removes the nominal error caused by approximate encoder dynamics and state reconstruction. An invertible component of the encoder provides the internal reconstruction needed to evaluate the KKL residual correction, making the synchronization manifold exactly invariant. A separate output map uses all latent coordinates while remaining an exact left inverse of the encoder for every value of its trainable parameters. We also established an ISS result for the corrected observer. In the Van der Pol and Lorenz examples, the corrected observers converged to the integration accuracy limit without measurement noise and generally reduced the noisy estimation errors relative to the released PINN--KKL implementations. 

We assessed the ISS conditions by numerical sampling rather than uniform
verification over the continuous domain. Future work will develop formal
certificates over the operating region, explore alternative invertible
encoders and contracting latent architectures, and evaluate the method on
broader classes of nonlinear and higher-dimensional systems.

\bibliographystyle{IEEEtran}
{\footnotesize
\bibliography{references}}
\end{document}